\documentclass[%
 reprint,
superscriptaddress,
 amsmath,amssymb,aps,prl,
 longbibliography,
floatfix,
]{revtex4-2}

\usepackage{graphicx}
\usepackage{dcolumn}
\usepackage{bm}
\usepackage{hyperref}
\usepackage[english]{babel}

\usepackage{xcolor} 

\begin{document}


\title{CW SRF Gun generating beam parameters sufficient for CW hard-X-ray FEL}

\author{Nikhil Bachhawat}
\affiliation{%
 Stony Brook University, Stony Brook, New York 11794, USA.
}%

\author{Vladimir N. Litvinenko}%
\affiliation{%
 Stony Brook University, Stony Brook, New York 11794, USA.
}%
\affiliation{%
 Brookhaven National Laboratory, Upton, New York 11973, USA.
}%
\author{Jean C. Brutus}
\affiliation{%
 Brookhaven National Laboratory, Upton, New York 11973, USA.
}%
\author{Luca Cultrera}
\affiliation{%
 Brookhaven National Laboratory, Upton, New York 11973, USA.
}%
\author{Patrick Inacker}
\affiliation{%
 Brookhaven National Laboratory, Upton, New York 11973, USA.
}%
\author{Yichao Jing}
\affiliation{%
 Stony Brook University, Stony Brook, New York 11794, USA.
}%
\affiliation{%
 Brookhaven National Laboratory, Upton, New York 11973, USA.
}%
\author{Jun Ma}
\affiliation{%
 Brookhaven National Laboratory, Upton, New York 11973, USA.
}%
\author{Igor Pinayev}
\affiliation{%
 Brookhaven National Laboratory, Upton, New York 11973, USA.
}%
\author{John Skaritka}
\affiliation{%
 Brookhaven National Laboratory, Upton, New York 11973, USA.
}%
\author{Gang Wang}
\affiliation{%
 Brookhaven National Laboratory, Upton, New York 11973, USA.
}%


\date{\today}

\begin{abstract}
SRF CW accelerator constructed for Coherent electron Cooling (CeC) Proof-of-principle (POP) experiment at Brookhaven National Laboratory has frequently demonstrated record parameters using 1.5 nC 350 ps long electron bunches, typically compressed to FWHM of 30 ps using ballistic compression. We report experimental demonstration of CW electron beam with parameters fully satisfying requirements for hard X-ray FEL and significantly exceeding those demonstrated by APEX LCLS II electron gun. This was achieved using a 10-year-old SRF gun with a modest accelerating gradient of $\sim$15 MV/m, a bunching cavity followed by ballistic compression to generate 100 pC, $\sim$15 ps FWHM electron bunches with a normalized slice emittance of $\sim$0.2 mm-mrad and a normalized projected emittance of $\sim$0.25 mm-mrad. Hence, in this paper, we present an alternative method for generating CW electron beams for hard-X-ray FELs using existing and proven accelerator technology. We present a description of the accelerator system settings, details of projected and slice emittance measurements as well as relevant beam dynamics simulations.
\end{abstract}

\maketitle


\textit{Introduction}---Superconducting RF (SRF) photoinjectors have emerged as reliable sources of high-brightness electron beams for CW linacs and X-ray free electron lasers (XFELs). Next-generation XFELs must be able to generate high-brightness, sub-micron emittance electron beams to demonstrate stable performance in CW mode~\cite{Petrushina_2020, Zhou_2023}.  It has been previously demonstrated that the Coherent electron Cooling (CeC) SRF CW gun can generate electron CW beams with record-low-emittances~\cite{Petrushina_2020}. 

However, it has been argued~\cite{Wang_2021} that the flat-top electron bunches generated by CeC SRF gun with a typical duration of 350-400 ps are too long for X-ray FEL applications. Instead, it is recommended that the CeC gun system generate electron bunches with $\sim$15-20 ps FWHM bunch length to demonstrate XFEL capabilities. While the main goal of our SRF accelerator is the generation of high-brightness, high-charge (0.8-1.5 nC) CW electron beams for the CeC Proof-of-Principle (POP) experiment, we were able to develop a low-charge operation mode in tandem. During Runs 24 and 25, we successfully demonstrated that the low-emittance beams generated by the CeC SRF gun can indeed be ballistically compressed to $\sim 15$~ps, which is the bunch length necessary for a hard-X-ray FEL injector, without a significant increase in transverse emittance. This demonstrates an alternative method of generating short low-emittance bunches necessary for hard-X-ray FELs.

In this Letter, we present the recent record-breaking performance of the CeC SRF gun and accelerator beamline for low charge electron beams (up to 100 pC), outperforming the current state-of-the-art LCLS II RF gun~\cite{Zhou_2023} (see Table~\ref{tab:cec_lcls_comparison}). We give an overview of the design and operation of the CeC accelerator and highlight the critical upgrades made to our system. This is followed by a description of our emittance studies using time-resolved diagnostics beam-line (TRDBL) and Dogleg. Finally, we present the experimental results from Run 24 and 25 for the low charge operation mode. Our results show that the CeC SRF gun satisfies the beam quality requirements for hard-X-ray CW FELs. Moreover, critical upgrades to low-emittance, high-QE NaKSb photocathodes, cathode transfer system and vacuum allowed for months-long operating lifetime that outperformed other RF guns~\cite{Wang_2021}.
\begin{table*}
    \caption{\label{tab:cec_lcls_comparison} Comparison of the Beam Quality between the CeC SRF Gun and the LCLS-II RF Gun ~\cite{Zhou_2023}}
    \begin{ruledtabular}
    \begin{tabular}{lcccc}
    \textbf{Parameter} & \textbf{Units} & \multicolumn{2}{c}{\textbf{LCLS II}} & \textbf{CeC} \\
    \cline{3-4}
     &  & projected & demonstrated &  \\
    \hline
    Gun Voltage & MeV & 0.75 & 0.65--0.75 & 1.25$^{a}$ \\
    Charge per bunch & pC & 100 & 20--100 & 50--20,000$^{a}$ \\
    Max. avg. beam current @ 100~pC & mA & 0.062 & 0.030 & 0.15$^{a}$ \\
    Transverse RMS norm. slice emittance @ 100~pC & mm-mrad & 0.4 & $\sim$0.5 & 0.2$^{a}$ \\
    Transverse RMS norm. projected emittance @ 100~pC & mm-mrad & - & - & 0.23$^{a}$ \\
    Longitudinal RMS slice emittance & keV$\cdot$ps & 3.3 & - & 0.7$^{b}$ \\
    Cathode QE & \% & 1 & $>$0.5 & $>$2$^{a}$ \\
    \end{tabular}
    \end{ruledtabular}
    \vspace{1ex}
    \raggedright
    $^{a}$Measured value \quad $^{b}$Extracted from simulations
\end{table*}

\begin{figure*}
\includegraphics[scale=0.62,trim=4 4 4 4,clip]{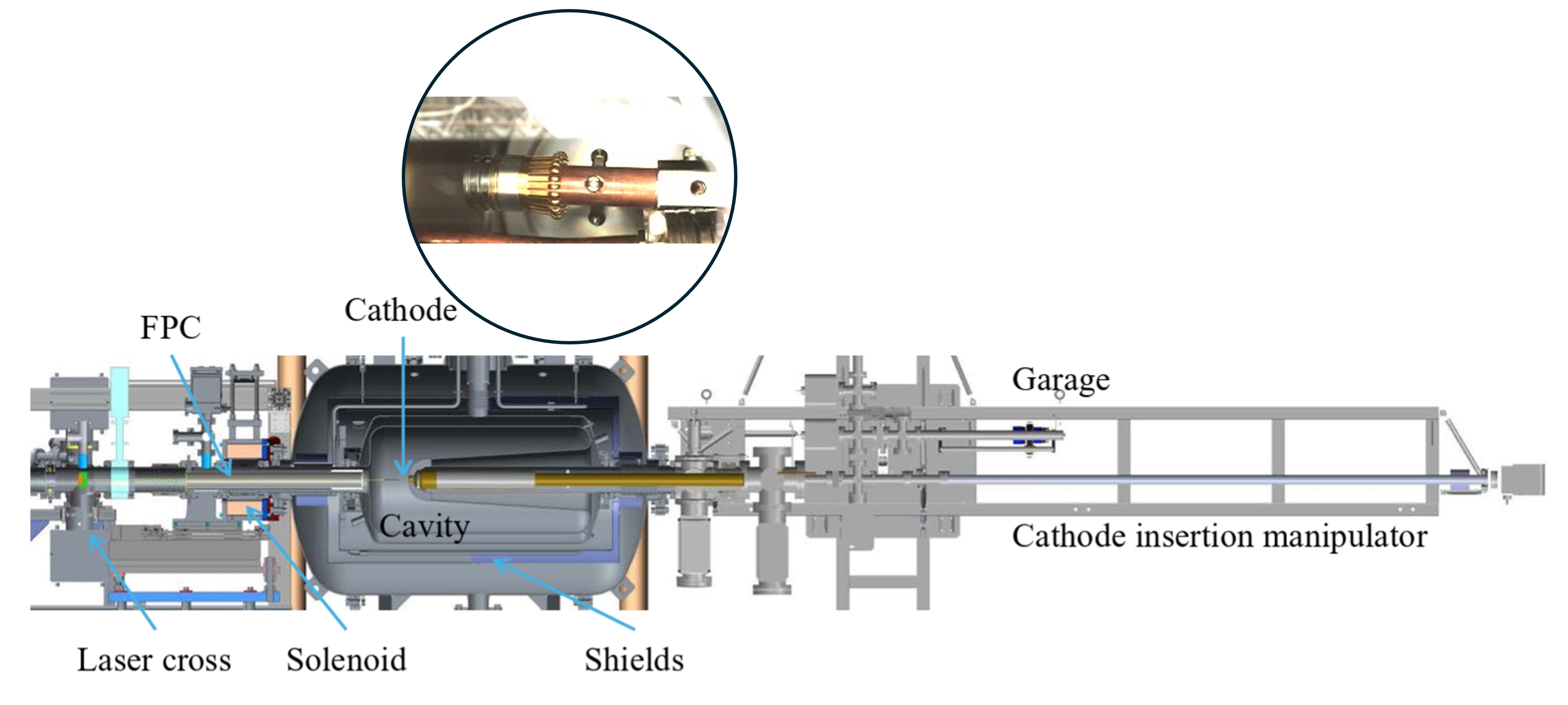}
\caption{\label{fig:srf_gun} CeC SRF CW gun and cathode transfer system.}
\end{figure*}

\textit{Experimental Setup}---Our 1.25 MeV SRF gun (see Fig.~\ref{fig:srf_gun}) has a 113 MHz quarter-wave resonator with a room temperature NaKSb photocathode inserted inside a cathode stalk serving as a half-wave RF choke. The cathode is illuminated by green (532 nm) laser pulses to generate electron bunches with a frequency of 78 kHz needed for the CeC POP experiment. The design bunch repetition rate of 78 kHz was chosen to match the revolution frequency of hadron beams in RHIC, which limits our average current to $\mathcal{O}$(100~\textmu A). While we have the capability to increase the repetition rate and operate with an $\mathcal{O}(1~ \mathrm{mA})$ beam, this program has very low priority. 

\begin{figure*}
\includegraphics[scale=0.53,trim=4 4 4 4,clip]{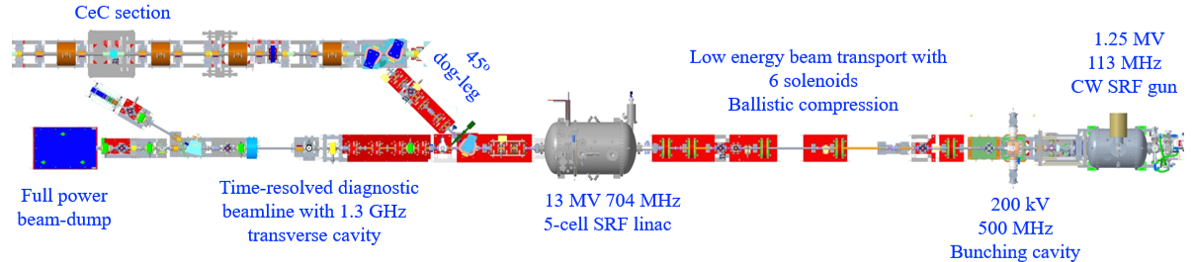}
\caption{\label{fig:cec_beamline} CeC SRF CW gun and cathode delivery system.}
\end{figure*}

While our SRF gun routinely operates at 1.25 MV CW, it is capable of operating at up to 1.5 MV CW (albeit with higher dark current and Li-He consumption) and up to 1.7 MV in pulsed mode. In case we observe a reduction in performance or a significant increase in radiation and dark current, we perform He conditioning of the cavity~\cite{Petrushina_2022}. Normally, we operate the CeC SRF gun with 350-400 ps FWHM 0.8-1.5 nC per bunch pulses to meet CeC requirements, but the gun has demonstrated total charge per bunch in excess of 20 nC. 

NaKSb photocathodes (with active areas from 9 mm to 14 mm in diameter) are deposited onto the surface of the Mo polished puck (20 mm in diameter). NaKSb cathodes have a desirable mean transverse energy (MTE) of 120-160 meV when illuminated with 532 nm green laser~\cite{Maxson_2015} for generation of low-emittance beams. However, natural degradation of QE of photocathode from as high as $4\%$ to $\sim0.3-0.4\%$ due to the cathode transfer process and gun operation, we expect the MTE to be $\sim$60 meV.

The pucks have two side grooves used for storage and in-vacuum transfers using forks. After deposition, three to five photocathodes are placed into an XUV vacuum suit (``garage"), transferred to the RHIC tunnel (where the SRF gun is installed) and placed into the UHV cathode transfer system with vertical and horizontal arms. Cathodes are taken from the garage and secured to the end-effector of the horizontal arm using spring-loaded balls. They are then inserted into the cathode stalk, whose room temperature inside the Li-He cooled 4~K environment is maintained by circulating water. The inner cylinder of the puck (10 mm in diameter) conducts both RF-induced electrical current and heat via an inner set of RF fingers to the end-effector and further through the outer RF fingers to the cathode stalk.

The stalk’s lateral position can be recessed with respect to the 4~K Nb nose of the SRF cavity. Such a recess, while reducing the field at the cathode surface, provides important RF focusing for generated electron beam. We found that for operation with 0.8-1.5 nC per bunch optimum recess is $\sim$10 mm~\cite{Petrushina_2019}. The QE lifetime of the photocathodes in our SRF gun is measured in months, and on a number of occasions we have operated with a single cathode for an entire RHIC run (i.e.~$\sim$6 months with $\mathrm{CsK_2Sb}$ photocathode). The observed photocathode performance was achieved through a series of targeted upgrades of the CeC system. These included the development of an XUV-scale, $10^{-12}$ torr vacuum suite and substantial improvements of vacuum in the gun cathode transfer system. 

The beam generated by the CeC SRF gun is then focused by a gun solenoid located 0.65 m downstream of the cathode. Subsequently, the beam energy is chirped in a room temperature 500 MHz bunching cavity, such that the beam undergoes ballistic compression in a low energy beam transport (LEBT) section equipped with five solenoids and a full set of dipole trims. Next, the compressed bunch is accelerated in a 13 MV 5-cell SRF linac operating at a frequency of 704 MHz. Finally, the accelerated beam is analyzed using a 45-degree dogleg and time-resolved diagnostics beamline (TRDBL), where we use quadrupole magnets downstream of the linac. A schematic of the CeC POP lattice is shown in Fig.~\ref{fig:cec_beamline}.

\textit{Emittance studies}---Several different methods of emittance measurements are used in tandem to evaluate beam quality throughout the CeC beamline. In the LEBT section, we predominantly use solenoid-based emittance measurements at two separate YAG screen locations - at YAG 1 between LEBT solenoids 1 and 2 and at YAG 2 between LEBT solenoids 3 and 4. At the YAG 2 location, we also have a pepper pot tool to measure projected emittance.

\begin{figure}
\includegraphics[scale=0.41,trim=4 4 4 4,clip]{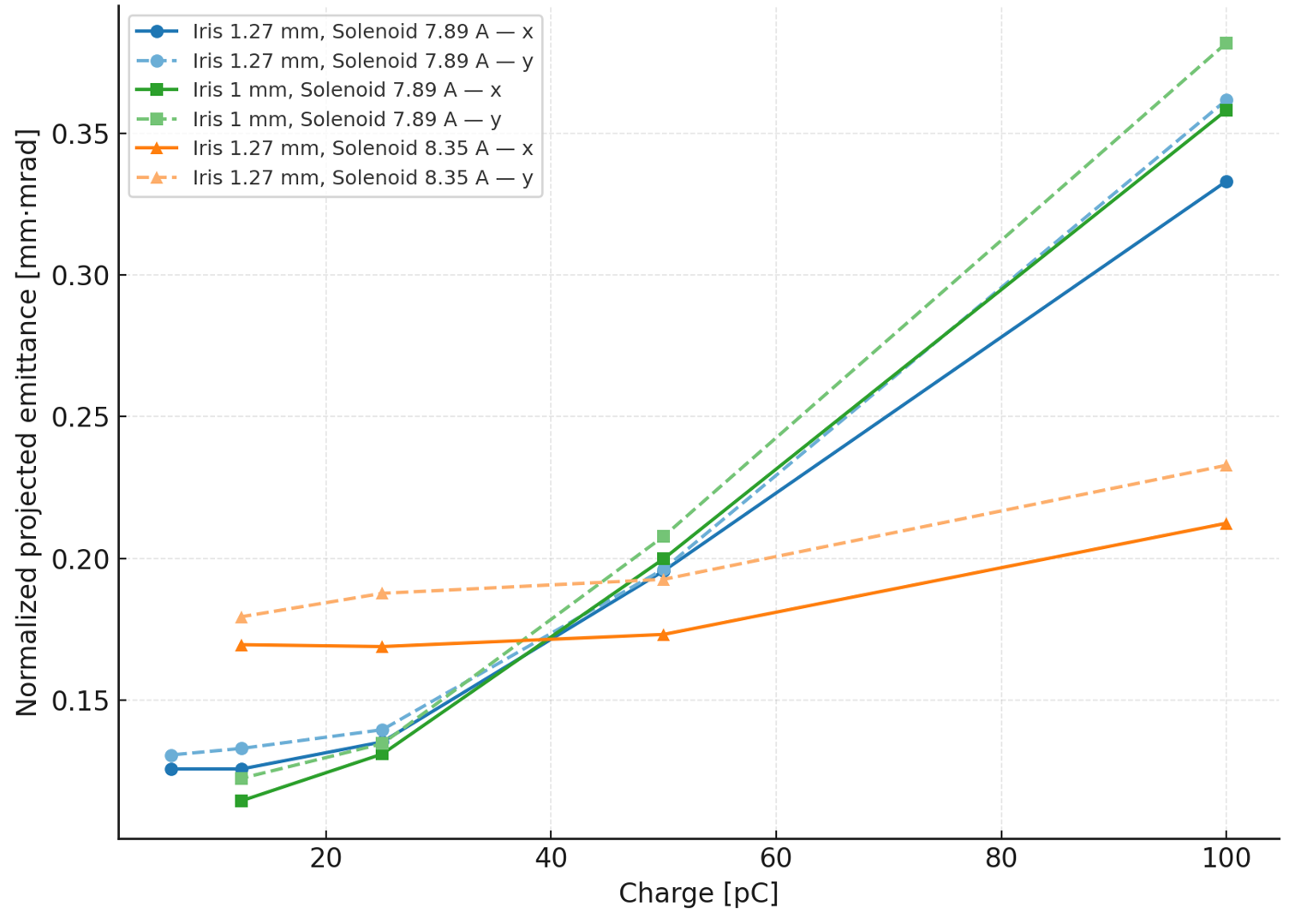}
\caption{\label{fig:yag1_emit_charge} Measured horizontal and vertical projected emittances for electron bunches with different total charge (6.25-100 pC). This is shown for three combinations of iris diameter on cathode and gun solenoid current, which give the best measured emittances at YAG 1 location.}
\end{figure}

The beam emittance is calculated from beam moments measured experimentally as a function of LEBT 1 and 3 solenoid currents. We show the transverse normalized projected emittance at YAG 1 location for beams with different total charges in Fig.~\ref{fig:yag1_emit_charge}. The emittances were measured for different combinations of iris diameter (spot size) on cathode and gun solenoid current which give the best measured emittances at YAG 1 location.

\begin{figure}
\includegraphics[scale=0.42,trim=4 4 4 4,clip]{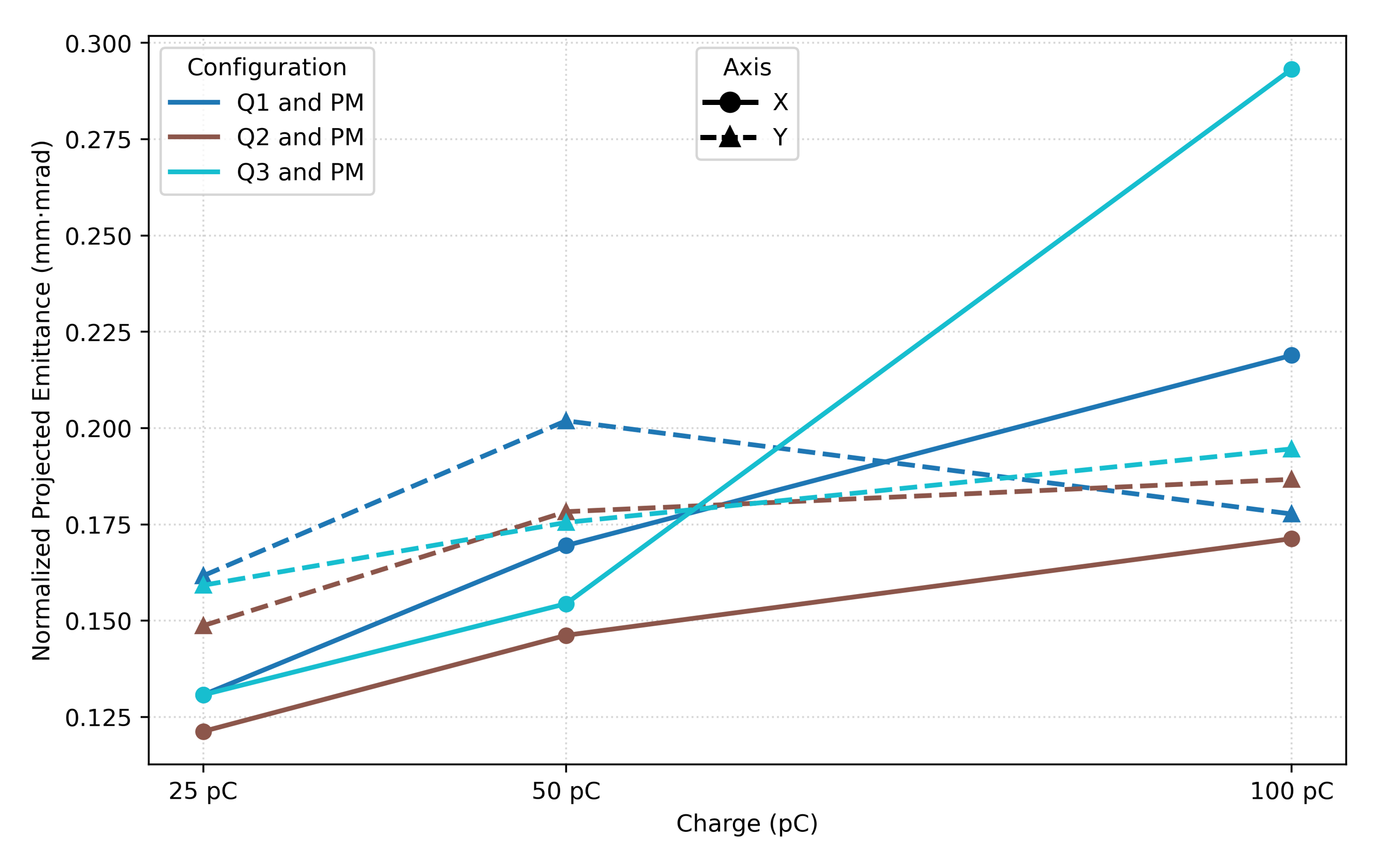}
\caption{\label{fig:TQ_TPM_charge_emit} Measured horizontal and vertical projected emittances for 25, 50 and 100 pC electron bunches using Triplet quadrupoles (Q1, Q2 and Q3) and Triplet PM in Dogleg section.}
\end{figure}

After the beam is ballistically compressed in the LEBT section, it is accelerated by a 704 MHz linac. Downstream from the linac, we can either steer the beam using a 45$^{\circ}$ bending dipole into the Dogleg section or allow the beam to continue propagation into the time-resolved diagnostics beamline (TRDBL), as shown in Fig.~\ref{fig:cec_beamline}. Under normal CeC operation, the Dogleg serves to prepare the beam for propagation through the amplifier and cooling section. In our low charge study, we make use of the three triplet quadrupoles along with a triplet position monitor (PM) as a diagnostic tool to measure transverse projected emittance immediately after the 704 MHz linac. We also show the comparison of the emittance measurements for different total beam charges and choice of triplet quadrupoles to perform the scan in Fig.~\ref{fig:TQ_TPM_charge_emit}. 
\begin{figure*}
\includegraphics[scale=0.52,trim=4 4 4 4,clip]{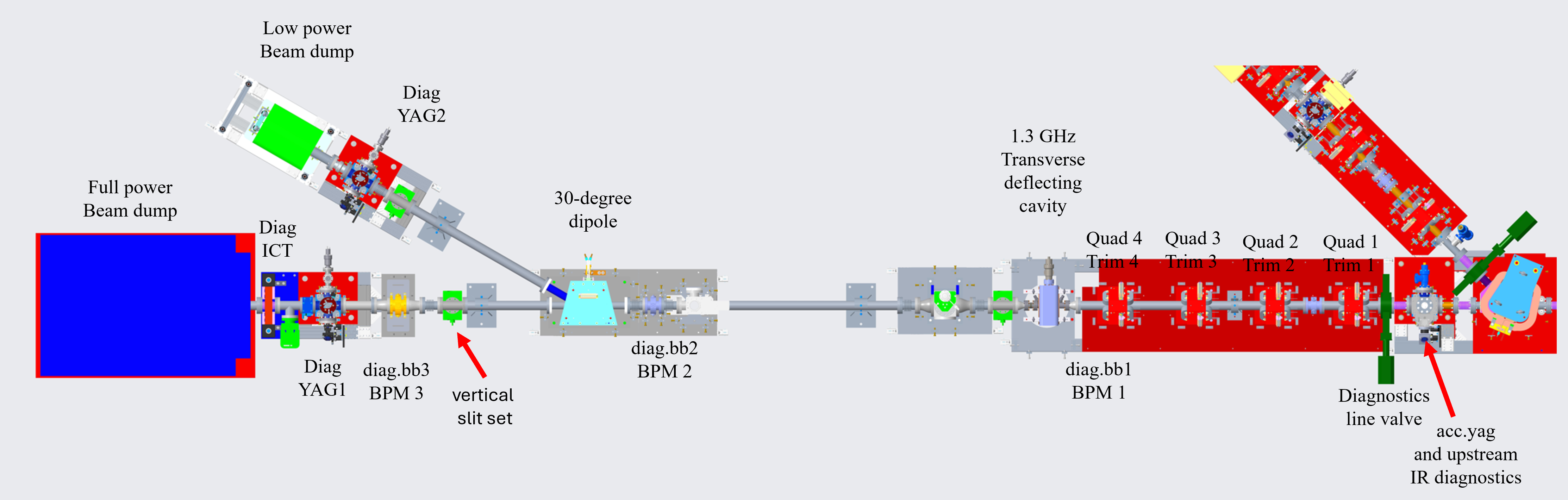}
\caption{\label{fig:cec_TRDBL} CeC time-resolved diagnostics beamline (TRDBL).}
\end{figure*}

When the bending dipole to the Dogleg section is turned off, the beam is delivered to the TRDBL. In the TRDBL, the electron beam goes through four quadrupoles (see Fig.~\ref{fig:cec_TRDBL}), where beam shape is optimized for beam diagnostics. The TRDBL has a 1.3 GHz transverse deflecting cavity where a time-dependent vertical kick is applied to the electron beam. This causes the different longitudinal slices of the electron bunch to move with different vertical velocities and hence, get separated downstream in the vertical plane~\cite{Wang_2021_2}. Analogous to the Dogleg section, diagnostic quadrupole and PM combinations can also be used in the TRDBL section to measure transverse projected emittances.

\begin{figure}
\includegraphics[scale=0.4,trim=4 4 4 4,clip]{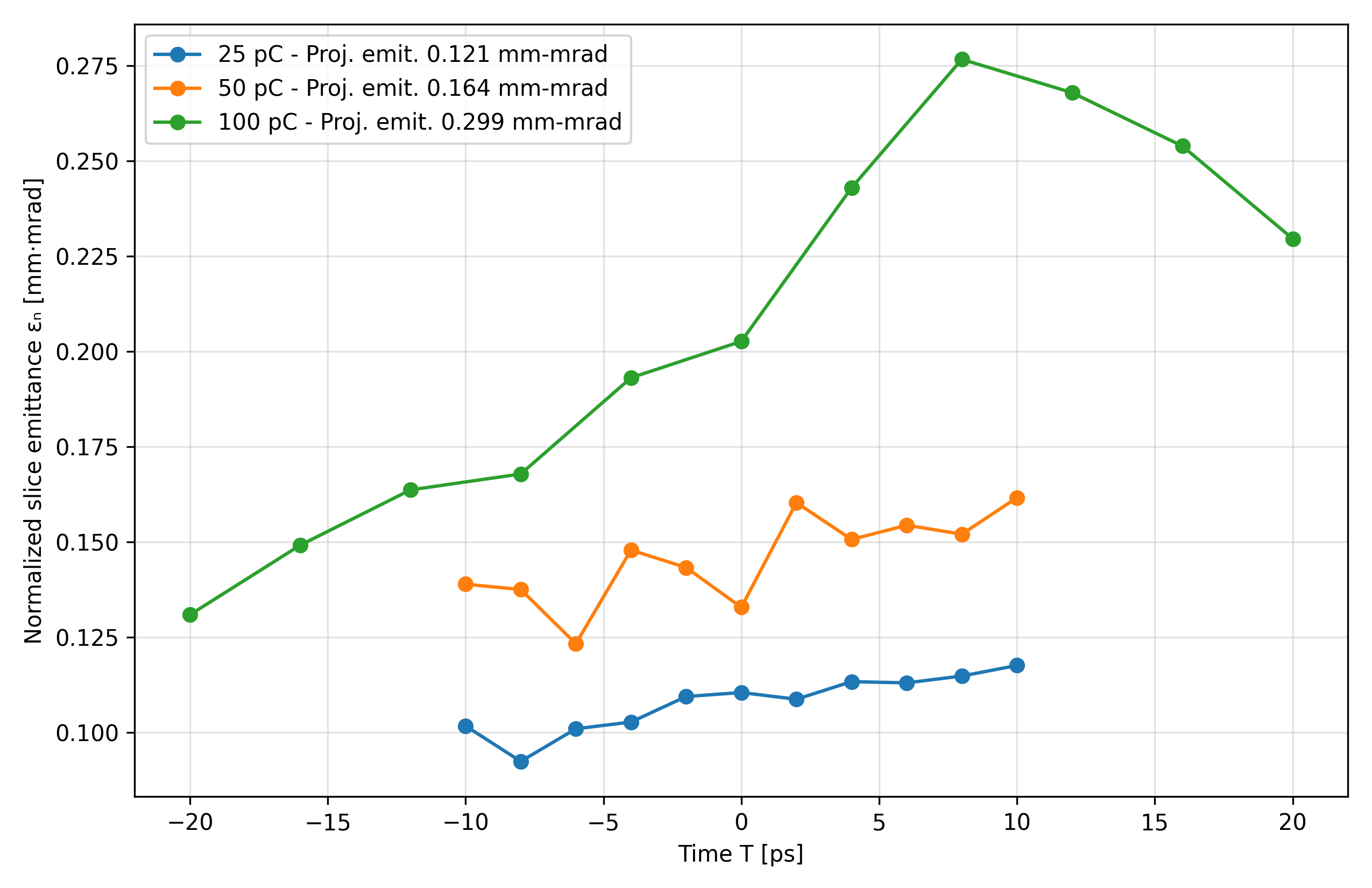}
\caption{\label{fig:slice_emit_charge} Measured normalized slice emittances for 25, 50 and 100 pC electron bunches using vertical slit, trim magnet current and Diagnostic YAG 1 in TRDBL section. We also report the projected normalized emittance calculated from the slice emittance measurements.}
\end{figure}

For slice emittance measurement, the bending magnet downstream of the deflecting cavity is turned off and we study the electron beam at the beam position monitor (BPM) in front of the high power beam dump, i.e. diagnostic YAG 1. We insert a vertical slit in front of the Diagnostic YAG 1 (see Fig.~\ref{fig:cec_TRDBL}) and then scan the beam passing through the slit by varying the trim magnet current. This allows us to measure the transverse phase-space and in turn, the beam moments, Twiss parameters and slice emittances. The slit-based slice emittance measurement has the additional benefit that it is robust in the space charge dominated regime as only a portion of the beam propagates through the slit. This is especially important under normal charge CeC operation (0.8-1.5 nC), but allows high quality measurements even in low charge operation. We show the measured slice emittances for different total beam charges (25, 50 and 100 pC) in Fig.~\ref{fig:slice_emit_charge}. 

\begin{figure}
\includegraphics[scale=0.4,trim=4 4 4 4,clip]{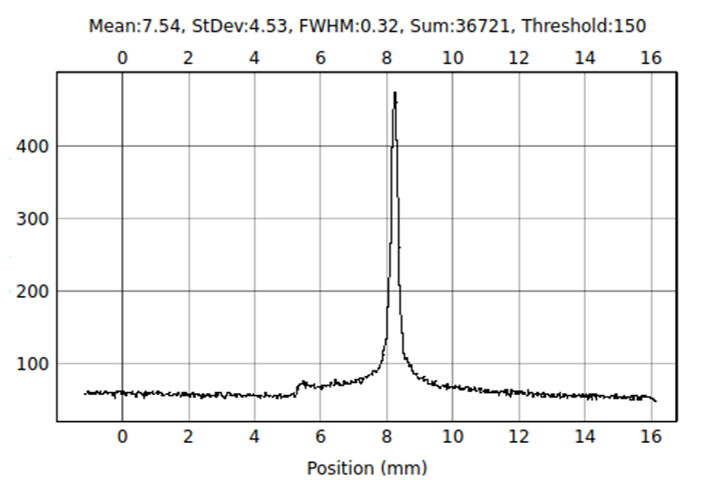}
\caption{\label{fig:energy_var} Snapshot in time of vertical projection of energy profile of 100~pC beam.}
\end{figure}

For longitudinal phase space imaging and slice energy spread measurement, the bending magnet is turned on and the electron beam reaches the beam profile monitor in front of the low power beam dump, i.e. diagnostic YAG 2. The 1.3 GHz deflecting cavity introduces a correlation between the longitudinal and transverse coordinates, allowing measurement of the bunch temporal profile directly using the diagnostic YAG 2 screen. We summarize the bunch length and peak current values achieved for 100 pC beam accelerated to 14.05 MeV and 9.5 MeV in Runs 24 and 25 respectively in Table~\ref{tab:long_beam_params}. 
\begin{table}
    \caption{\label{tab:long_beam_params} Bunch length and peak current values for 100 pC beam from Run 24 (14.05 MeV KE) and 25 (9.5 MeV KE)}
    \begin{ruledtabular}
    \begin{tabular}{cccc}
       \textbf{KE} & \multicolumn{2}{c}{\textbf{Bunch length}} & \textbf{Peak current} \\ 
       \cline{2-3}
        \textbf{(MeV)}& RMS & FWHM & \textbf{(A)} \\
        \hline
        14.05 & 7.3 & 14.4 & 3.5 \\
        9.5 & 10.2 & 17.7 & 4.9 \\
    \end{tabular}
    \end{ruledtabular}
    \raggedright
\end{table}
We were able to successfully demonstrate that the CeC lattice can ballistically compress 100~pC electron bunches to 14-17 ps FWHM, with $2.5\times 10^{-4}$ FWHM relative energy spread ($1\times 10^{-4}$ RMS spread). A time snapshot of the vertical projection of the energy profile is shown in Fig.~\ref{fig:energy_var}.

\textit{Conclusion}---We experimentally show that our CeC SRF gun can generate electron bunches with duration $\sim$15~ps FWHM, less than 0.25 mm-mrad normalized projected emittances and $\sim$0.2 mm-mrad normalized slice emittance. This result confirms that the low transverse emittance beams generated by the CeC SRF gun~\cite{Petrushina_2020} can also be ballistically compressed to the bunch length required for a hard-X-ray FEL injector, without any significant increase in transverse emittance. In Table~\ref{tab:cec_lcls_comparison}, we provide a comparison of the CeC SRF gun with the demonstrated and projected performance of the state-of-the–art LCLS II X-ray FEL gun~\cite{Zhou_2023}. These record-breaking results confirm that our 10-year-old CeC SRF gun establishes an alternative method to generate short, low-emittance beams of sufficient quality for CW hard-X-ray FEL applications.
\\

\textit{Acknowledgments}---We would like to acknowledge assistance provided by K. Shih and J. Qiang with IMPACT-T simulations. This research also used computer resources at the National Energy Research Scientific Computing Center (NERSC).


\bibliography{low_charge_paper}

\end{document}